\documentclass{PoS}

\pdfoutput=1

\usepackage{amsmath}
\usepackage{subfig}
\usepackage{multirow}

\let\oldbibliography\thebibliography
\renewcommand{\thebibliography}[1]{%
  \vspace{-1.1ex}
  \oldbibliography{#1}%
  \vspace{-2ex}
  \setlength{\itemsep}{0pt}%
}

\newcommand{\pbp}{\langle\bar\psi\psi\rangle}
\newcommand{\mr}[2]{\multirow{#1}{*}{#2}}

\newcommand{\ls}[1]{#1^3\!\!\times\!32}

\newcommand{\eI}[1]{\phantom{(}\llap[#1\rlap]\phantom{)}}

\title{Lattice QCD with 12 Degenerate Quark Flavors}

\ShortTitle{Lattice QCD with 12 Degenerate Quark Flavors}

\author{\speaker{Xiao-Yong Jin}\\
        Department of Physics, Columbia University, New York, NY 10027, USA\\
        E-mail: \email{xj2106@columbia.edu}}

\author{Robert D. Mawhinney\\
        Department of Physics, Columbia University, New York, NY 10027, USA\\
        E-mail: \email{rdm@physics.columbia.edu}}

\abstract{

  We report on new data from additional zero temperature simulations
  of QCD with 12 flavors.  This is a continuation of previous studies
  using the DBW2 gauge action and naive staggered fermions.  With the
  use of the force gradient integrator and a multiple-quark-mass
  preconditioned HMC, we have done simulations with input quark masses
  from $m_q=0.003$ to $m_q=0.008$.  We have observed a metastable,
  first order, bulk transition that occurs at small input quark
  masses.  As the quark mass increases, this first order bulk
  transition ends at a second order critical point, and, for still
  heavier quark masses, becomes the cross-over we have previously
  reported.  We present measurements of hadron masses, decay constants
  and other low energy observables in the small quark mass region on
  the weak coupling side of the bulk transition.  Our results show
  that the behavior of the system is still consistent with
  spontaneously broken chiral symmetry.  We also discuss a preliminary
  investigation into the behavior of the bulk transition itself.  We
  have found that, as the system approaches the second order critical
  end point, the scalar singlet meson becomes lighter.  Thus it
  appears that the critical endpoint corresponds to a continuum limit
  theory only involving scalars and, following known triviality
  arguments, this is likely a free field theory.  The presence of this
  critical endpoint could influence scaling of lattice observables in
  the conventional continuum limit.

}

\FullConference{The XXIX International Symposium on Lattice Field Theory - Lattice 2011\\
July 10-16, 2011\\
Squaw Valley, Lake Tahoe, California}

\begin{document}

\vspace{-1ex}
\section{Introduction}
\vspace{-2ex}

The zero temperature phase of QCD with many degenerate massless quark
flavors has been a heated topic for a number of years.  It is
important to establish if an infrared fixed point is dynamically
generated by a non-abelian gauge theory.  It enables all the
theoretical possibilities with a conformal symmetry (if the fixed
point exists) or a walking coupling constant (if the number of flavors
is less than required for a fixed point), which leads to some exciting
theories as candidates for beyond the standard model physics.

Lattice methods are indispensable for determining the critical number
of quark flavors required for the fixed point, as it is a
non-perturbative phenomenon.  Various groups are working on the
subject.  As the situation is new and unlike QCD in nature, we need to
be careful in our interpretations of the numerical results from
lattice calculations.

We have been using the DBW2 gauge and the naive staggered fermion
action to simulate the SU(3) gauge theory with 8 and 12 flavors.  In
this report, we will discuss our investigation into a first order bulk
transition separating strong and weak couplings, for small quark
masses, and the second order critical point, where one expects the
bulk transition to end, as the quark mass is increased.  We argue that
such a critical point (a lattice artifact) can alter the scaling
behavior of the hadronic observables, if simulations are done in
regions of the parameter space where the critical point has influence,
and may have an impact on the running of the coupling constant
directly measured on the lattice.

\vspace{-1ex}
\section{First Order Bulk Transition}
\vspace{-2ex}

\begin{figure}
  \vspace{-3ex}
  \centering
  \subfloat[8 flavors, $m_q=0.015$]{\label{fig:8f_wilson}
    \includegraphics[width=0.49\textwidth]{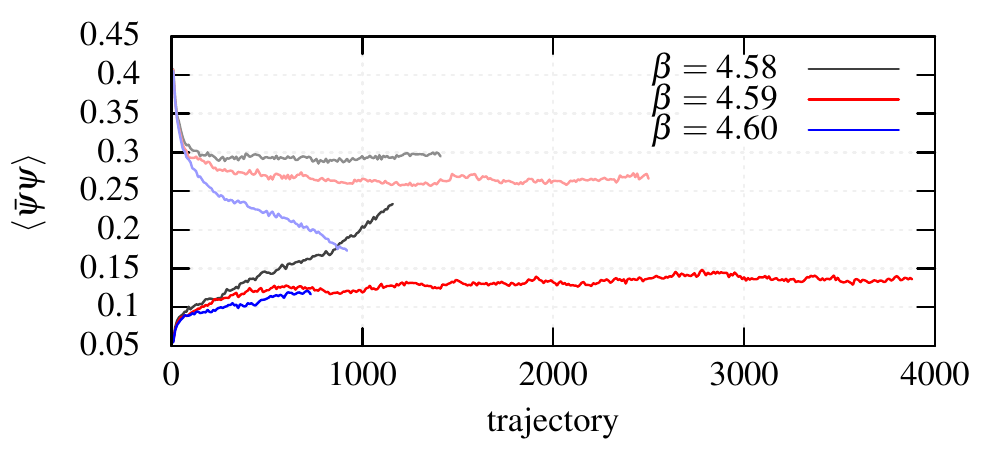}}
  \subfloat[12 flavors, $m_q=0.01$]{\label{fig:12f_wilson}
    \includegraphics[width=0.49\textwidth]{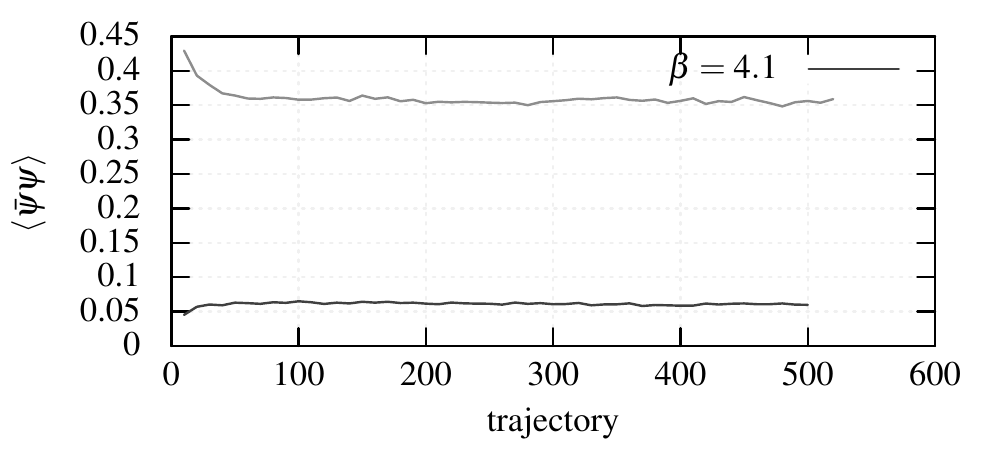}}
  \caption{The evolution of $\pbp$ with 8 quark flavors (left) and 12
    quark flavors (right), containing ordered (bottom curves) and
    disordered (top curves) starts, both with a lattice size of
    $16^3\times32$.}
\end{figure}

A bulk transition has been found and studied in an earlier
paper\cite{Brown:1992fz}, for lattice QCD with 8 quark flavors, using
the Wilson gauge action and naive staggered fermions, with the inexact
R algorithm.  The bulk transition was interpreted as a lattice
artifact.  It is independent of the lattice volume and temperature,
and it is not associated with a deconfinement transition or chiral
symmetry restoration.  Recently, using the exact RHMC algorithm, we
have rediscovered the bulk transition with the same gauge and fermion
action as in the earlier study.  We found it at $\beta=4.59$ and
$m_q=0.015$ with a lattice size of $16^3\times32$, shown in
figure~\ref{fig:8f_wilson}.  Such a bulk transition, however, did not
occur within the parameter space we have surveyed in our recent
studies~\cite{Jin:2008rc,Jin:2009mc,Jin:2010vm} with the DBW2 gauge
and naive staggered fermion action.

One might expect the lattice artifact bulk transition to become
stronger with 12 quark flavors.  The increase in the number of flavors
makes the renormalized coupling run more slowly, and a higher lattice
cutoff is required to approximate the continuum.  In other words, at a
comparable lattice spacing, the lattice gauge field becomes rougher
with increasing number of flavors.  The lattice consequently becomes a
poorer approximation of the continuum.  Observation of a bulk
transition is mentioned in a study of a lattice Schrodinger functional
calculation~\cite{Appelquist:2009ty}.  For 12 flavor QCD, with the
same Wilson gauge and naive staggered fermion action, we have located
the position of a first order bulk transition at $\beta=4.1$ and
$m_q=0.01$ with a lattice size of $16^3\times32$ and the exact HMC
algorithm, as shown in figure~\ref{fig:12f_wilson}.

In our recent study of QCD with 8 and 12 flavors~\cite{Jin:2009mc}
with the DBW2 gauge and the naive staggered fermion action, it is
intriguing to see that there is no meta-stability signal, but only a
quark mass dependent, quick cross-over region.  Nevertheless, the
strong dependence on the quark mass suggests that a first order bulk
transition may occur with smaller quark masses than we had been able
to simulate with.  Simulations with smaller quark masses became viable
with algorithmic improvements.  Using the force gradient integrator,
which is accurate up to the $4^\text{th}$ order, combined with
multiple Hasenbusch mass preconditioning, we were able to use the
exact HMC algorithm to explore the smaller quark mass region.

\begin{figure}
  \vspace{-3ex}
  \centering
  \subfloat[$\pbp$]
  {\includegraphics[width=0.58\textwidth]{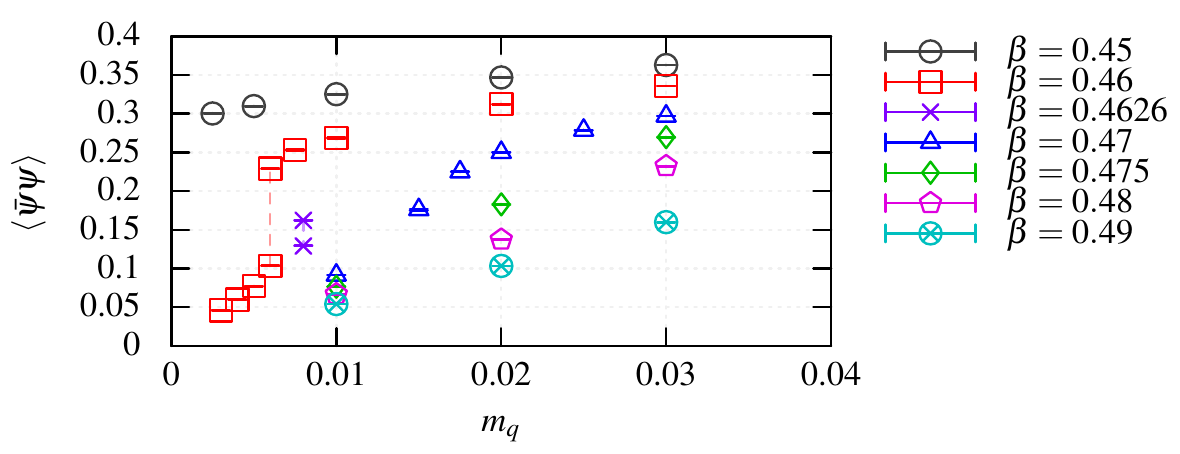}}
  \subfloat[$f_\pi$]
  {\includegraphics[width=0.41\textwidth]{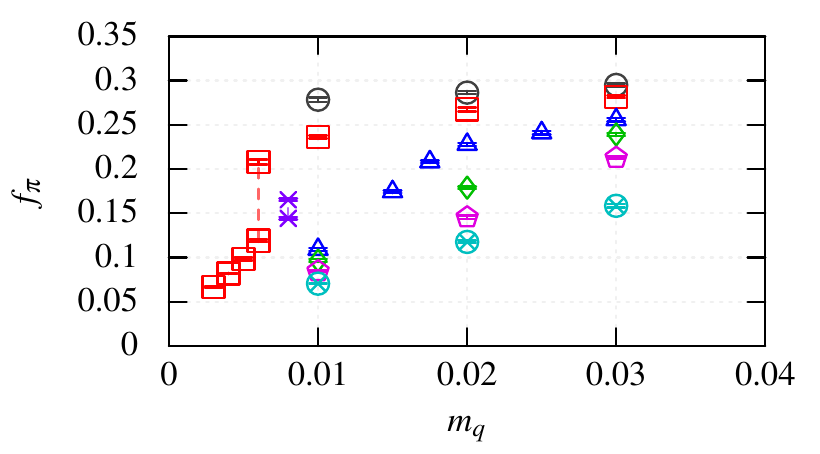}}
  \caption{\label{fig:12f_pbp_fpi} $\pbp$ (left) and $f_\pi$ (right)
    with 12 flavors.  Dashed lines indicates the meta-stable signal
    from the first order bulk transition, seen for both $\beta=0.46$
    and $0.4626$.}
\end{figure}

\begin{figure}
  \vspace{-1.5ex}
  \centering
  \subfloat[$\beta=0.46$, $m_q = 0.006$]
  {\includegraphics[width=0.49\textwidth]{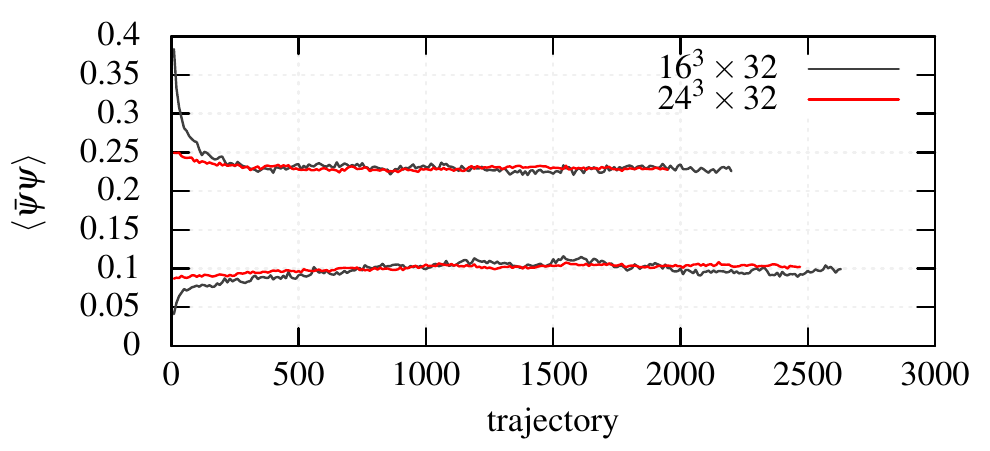}}
  \subfloat[$\beta=0.4626$, $m_q = 0.008$]
  {\includegraphics[width=0.49\textwidth]{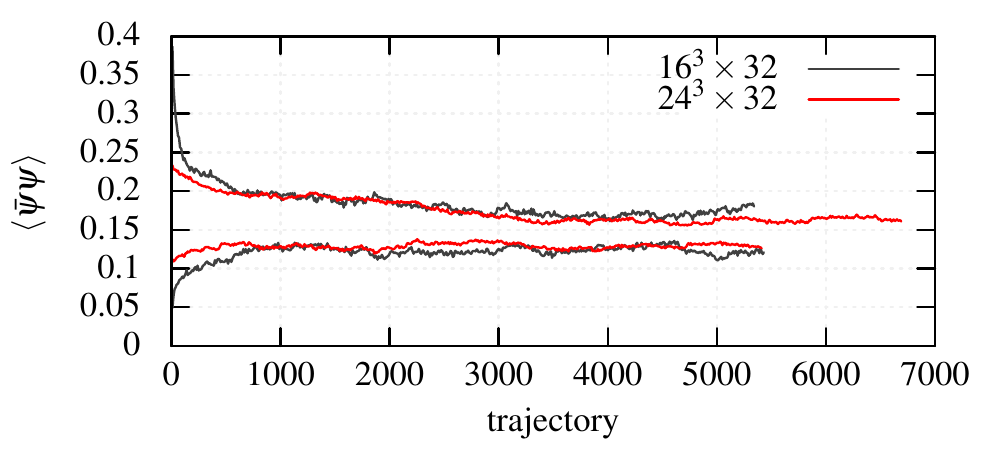}}
  \caption{\label{fig:12f_bulk} The evolution of $\pbp$ with DBW2
    gauge and naive staggered fermion action.  Two combinations of
    $\beta$ and $m_q$ show meta-stable evolutions with separate curves
    for ordered-starts (bottom curves) and disordered-starts (top
    curves).  The metastability is present for two different volumes.}
\end{figure}

A first order bulk transition was indeed found with 12 quark flavors
at smaller quark masses, using the DBW2 gauge and naive staggered
fermion action.  The location of the bulk transition is shown in
figure~\ref{fig:12f_pbp_fpi}.  It shows the first order, meta-stable
points (those have two points at the same $\beta$ and $m_q$), along
with other values\footnote{In this paper, $f_\pi$ is normalized to be
  about $93\,\mathrm{MeV}$ for QCD in physical units.  Values of
  $f_\pi$ in our previous reports~\cite{Jin:2009mc,Jin:2010vm} are off
  by a factor of 2, because of a misunderstanding of the convention of
  the CPS.} from our earlier simulations.  In the evolution of $\pbp$,
shown in figure~\ref{fig:12f_bulk}, we have found meta-stable signals
at two different couplings and quark mass combinations.  The
meta-stable evolution lasted for a few thousand trajectories.  This
first order bulk transition has no visible volume dependence, as seen
from comparing two different volumes, $16^3\times32$ and
$24^3\times32$.  Therefore, it is unlikely to be related to the
thermal transition.  In addition, as discussed
previously~\cite{Jin:2009mc}, there is no visible change in the values
of Polyakov loops, and there are clearly non-zero string tensions
found in the inter-quark potentials at both sides of the transition.
We can conclude that this bulk transition is not a deconfinement
transition.  On the other hand, the chiral symmetry is spontaneously
broken on both sides of the bulk transition, as argued in our previous
report~\cite{Jin:2009mc}.  The signal from all the hadronic
observables is clear and well justified at the strong coupling side of
the transition.  On the weak coupling side, the linearity of $m_\pi^2$
versus $m_q$ suggests the existence of a dimensionful quantity $B$, as
in leading order chiral perturbation theory where $m_\pi^2 = 2Bm_q$.
Consequently, with a non-zero $f_\pi$ in the chiral limit, the GMOR
relation, $f_\pi^2m_\pi^2=2m\pbp$, predicts a small but non-zero
chiral condensate in the chiral limit, which is consistent with out
data.  This argues that the weak coupling side of the bulk transition
also has the chiral symmetry spontaneously broken.  More of the
spectrum results will be shown in the next section.  In conclusion,
the bulk transition does not restore chiral symmetry from the data we
have collected.

Our simulations suggest the following phase structure.  There is a
first order bulk transition line in the $\beta$ versus $m_q$ plane, at
smaller masses, $m_q<m_q^C$.  At larger masses, $m_q>m_q^C$, the bulk
transition becomes a mass dependent rapid cross-over.  In addition to
the usual Gaussian fixed point at $\beta=\infty$ and $m_q=0$, where
the lattice gauge field approaches the continuum limit asymptotically,
the results from our study of the system with 12 quark flavors suggest
that there is an additional critical point at $\beta=\beta^C$ and
$m_q=m_q^C$.  It is the end point of the first order bulk transition,
a second order phase transition point, where certain correlation
lengths diverge.  It also represents a continuum theory.

The position of the critical point depends on the lattice action.  In
our case, the change from the unimproved Wilson gauge to the DBW2
gauge action moves the critical point to smaller quark masses in
lattice units.

With 8 flavors, it is also possible that the bulk transition does not
go away completely, after we have switched to DBW2 from unimproved
Wilson gauge.  The bulk transition and the second order critical point
may persist at a smaller quark masses away from the quark mass region
we have analyzed.  This possibility, however, does not invalidate our
conclusions about the phase of the QCD system with 8 flavors.  The
spontaneous breaking of chiral symmetry and the asymptotic freedom
holds at both side of the bulk transition.

Nevertheless, it is plausible that this additional critical point
(different from $\beta=\infty$) is capable of affecting the scaling of
the hadronic variables and also changing the running of the coupling
constant directly measured on the lattice.  More careful and thorough
studies are needed to investigate the effect on the running of the
coupling constant.  Yet, continuing our recent studies, we are able to
identify, detailed in the next section, the effect of this bulk
transition critical point in the meson spectrum.

\vspace{-1ex}
\section{The Meson Spectrum and the Scalar Singlet Meson}
\vspace{-2ex}

\begin{table}
  \vspace{-3ex}
  \footnotesize
  \centering
  \begin{tabular}[c]{llllllll}
    $\beta$        & $m_q$         & Size      & $f_\pi$     & $m_\pi$     & $m_{\sigma^c}$ & $m_\rho$      & $m_{a_1}$    \\
    \hline\hline
    \mr{7}{0.46}   & \mr{2}{0.003} & $\ls{24}$ & 0.06782(74) & 0.19462(42) & 0.3429(37)     & 0.4458(41)    & 0.566(18)    \\
                   &               & $\ls{32}$ & 0.06661(54) & 0.19370(33) & 0.3418(25)     & 0.4338(34)    & 0.585(11)    \\
    \cline{2-8}
                   & 0.004         & $\ls{24}$ & 0.08192(71) & 0.21633(53) & 0.3945(43)     & 0.5016(62)    & 0.638(27)    \\
    \cline{2-8}
                   & \mr{2}{0.005} & $\ls{16}$ & 0.0947(12)  & 0.23519(91) & 0.436(15)      & 0.589(10)     & 0.96(10)     \\
                   &               & $\ls{24}$ & 0.0983(16)  & 0.23278(53) & 0.4811(99)     & 0.6193(97)    & 0.847(83)    \\
    \cline{2-8}
                   & 0.006$^O$     & $\ls{24}$ & 0.1191(14)  & 0.24368(28) & 0.565(16)      & 0.756(12)     & 0.995\eI{94} \\
                   & 0.006$^D$     & $\ls{24}$ & 0.2080(28)  & 0.21068(14) & 0.882(97)      & 1.151\eI{13}  & 1.61\eI{51}  \\
    \hline
    \mr{2}{0.4626} & 0.008$^O$     & $\ls{24}$ & 0.1442(15)  & 0.26973(18) & 0.6667\eI{86}  & 0.8444\eI{49} & 1.099\eI{37} \\
                   & 0.008$^D$     & $\ls{24}$   & 0.1652(16)  & 0.25926(16) & 0.773(16)      & 0.986\eI{16}  & 1.226\eI{52} \\
  \end{tabular}
  \caption{\label{tab:spect} Some of the meson masses measured in the
    small quark mass region.}
\end{table}

Results for some meson masses are shown in table~\ref{tab:spect} for
$\beta=0.46$ and $\beta=0.4626$, respectively, at $m_q \le m_q^C$.
Wall2Z sources are used for the propagators, except for $f_\pi$, which
we measured from point to point propagators.  In the table, the lable
$^O$ or $^D$ on the quark mass indicates an ordered (weak coupling
side) or disordered (strong coupling side) start, respectively.  Most
of the errors are statistical, except those enclosed by angular
brackets have been increased to include a systematic variation caused
by varying the fitting range.

\begin{figure}
  \vspace{-1.5ex}
  \centering
  \subfloat[$m_\pi^2$ versus $m_q$]
  {\label{fig:12f_mpi_below_bulk}\includegraphics[width=0.49\textwidth]{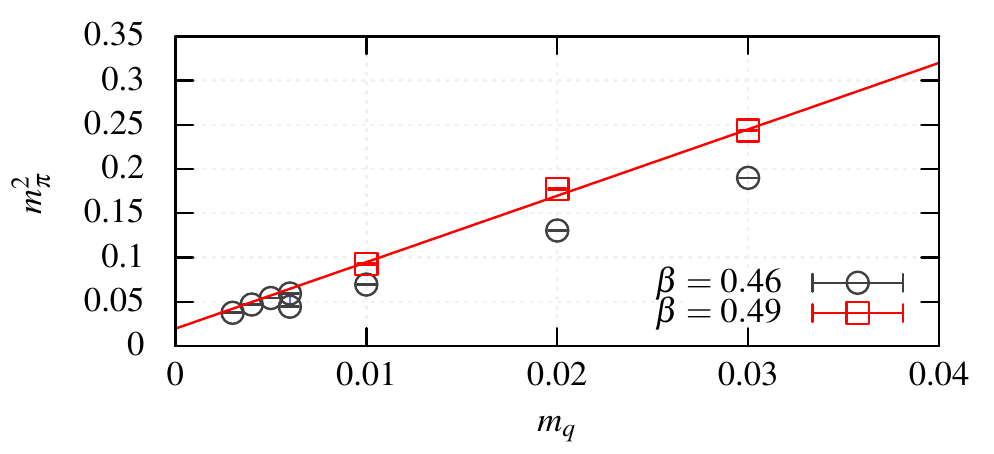}}
  \subfloat[$m_\rho$ and $m_{a_1}$ versus $m_q$]
  {\label{fig:12f_mrhoa1_below_bulk}\includegraphics[width=0.49\textwidth]{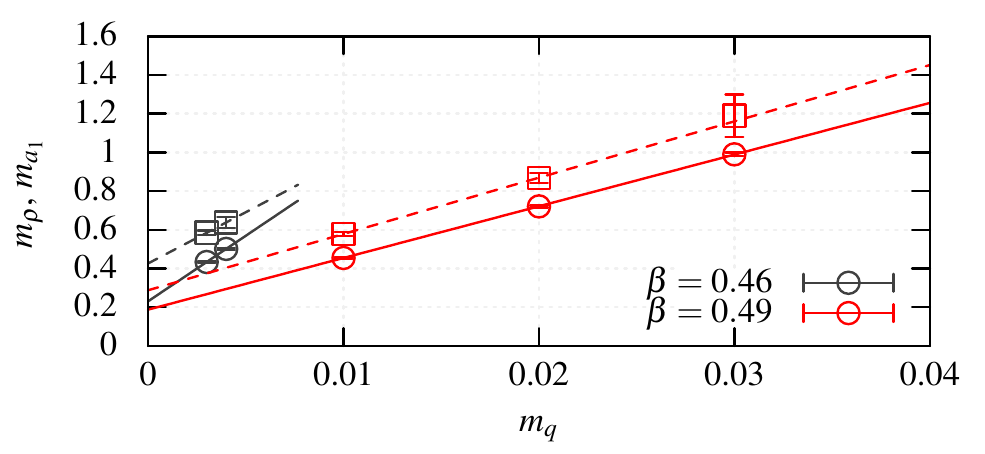}}
  \caption{Masses of the $\pi$, $\rho$ and $a_1$ mesons versus input
    quark mass, $m_q$, at $\beta=0.46$ and $\beta=0.49$.}
\end{figure}

Figure~\ref{fig:12f_mpi_below_bulk} shows that, on both sides of the
bulk transition, there is the profound linearity of $m_\pi^2 = 2Bm_q$.
This suggests that the chiral symmetry is spontaneously broken on both
sides of the bulk transition.  It is confirmed in the mass splitting
between $\rho$ and $a_1$ mesons, as depicted in
figure~\ref{fig:12f_mrhoa1_below_bulk}.

There is still one thing left unexplained.  It is well known that the
correlation length diverges as a system approaches a second order
critical point, which means some mass in the system with 12 flavors
should vanish at the critical point.  We can see from
table~\ref{tab:spect} that, while the system goes from strong to weak
coupling through the bulk transition, in lattice units, the masses of
$\sigma$, $\rho$ and $a_1$ mesons drop, and the pion mass increases.
This means that, going from strong to weak coupling, the lattice
constant or the scale of the system decreases.  However, none of the
masses here appears to be vanishing toward the critical point, from
$\beta=0.46$, $m_q=0.006$, to $\beta=0.4626$, $m_q=0.008$.

With further study, we have found that it is the scalar singlet meson
whose mass appears to be vanishing near the critical point.  Remember
that the $\sigma$ meson masses shown in table~\ref{tab:spect} are only
from the connected propagators.  To measure the disconnected scalar
propagators, we employed random Z2 volume sources.  We have found
marginal advantages with the Z2 against the U1 random source.  We used
40 random sources for each gauge configurations to estimate the quark
loops with $G(\tau) = \langle m \phi^\dag\phi \rangle$, as proposed by
the authors of the paper~\cite{Venkataraman:1997xi}, where $M \phi =
\eta$ with $\eta$ the random source and $M$ the fermion matrix.  The
singlet propagators, then, can be assembled as
\begin{equation}
  \label{eq:1}
  C(t) = \Big\langle
  \frac{N_f}{4} \langle G(\tau)G(\tau+t) \rangle_\tau
  - C^c(t)
  \Big\rangle
  - \frac{N_f}{4}
  \Big\langle \langle G(\tau) \rangle_\tau \Big\rangle^2
  \longrightarrow A \left(e^{-mt} + e^{-m(n_\tau-t)}\right)\,,
\end{equation}
where $C^c(t)$ is the connected point to point propagator,
$\langle\cdots\rangle_\tau$ is average over time slices, and
$\langle\cdots\rangle$ is ensemble average.  In the equation, a factor
of $N_f$ is removed from the usual definition, and the factor of $4$
comes from the staggered Dirac operator.  We used a single mass state
to fit the propagator to obtain the singlet mass, as the opposite
parity state is numerically invisible to us.

\begin{figure}
  \vspace{-3ex}
  \centering
  \subfloat[$\beta=0.46$, $m_q=0.006$]
  {\label{fig:scalar_em_b046}\includegraphics[width=0.49\textwidth]{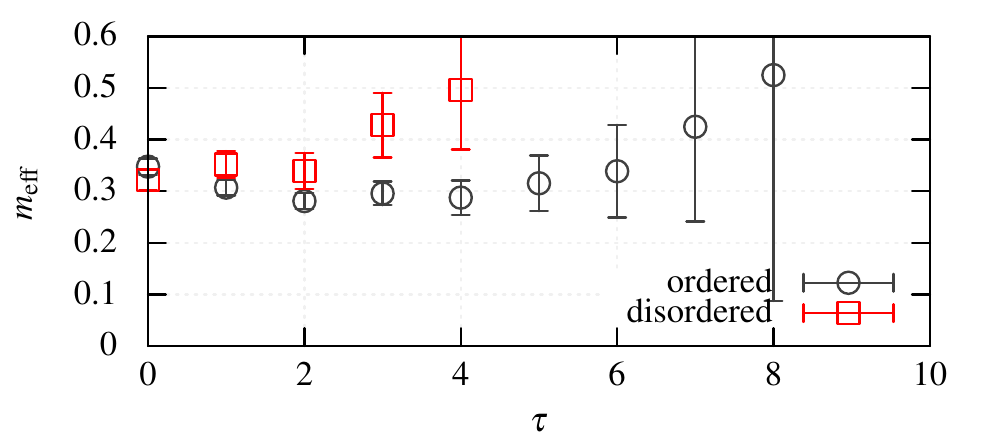}}
  \subfloat[$\beta=0.4626$, $m_q=0.008$]
  {\includegraphics[width=0.49\textwidth]{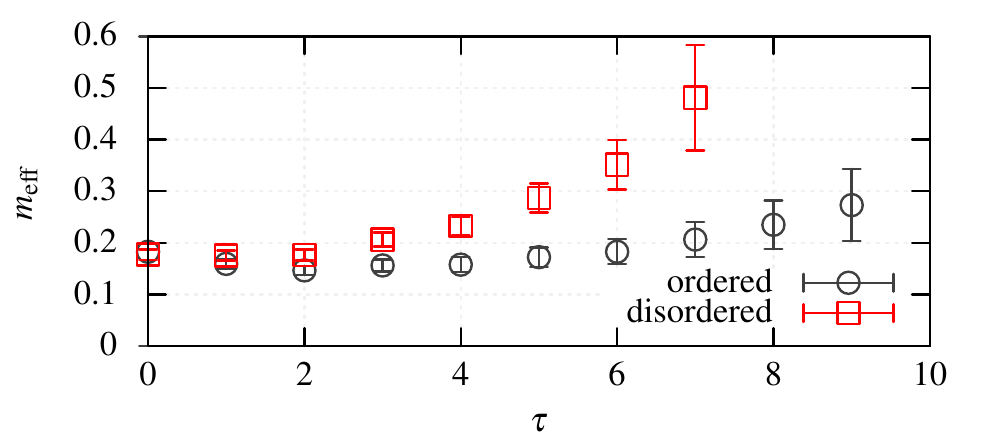}}
  \caption{\label{fig:scalar_em}Effective masses of the scalar singlet
    meson.  Gauge configurations with ordered and disordered starts
    respectively correspond to weak and strong coupling sides of the
    transition.}
\end{figure}

Figure~\ref{fig:scalar_em} shows the effective masses of the scalar
singlet meson at the bulk transition.  The value of $m_\text{eff}$ at
$\tau$ is obtained from $C(\tau)$ and $C(\tau+1)$.  It shows a bizarre
behavior for all four propagators.  When the temporal separation
becomes larger, the effective mass starts increasing.  However, it is
understandable.  The singlet propagator becomes noisier and more
difficult to measure with increasing temporal separations.  The
expected behavior is demonstrated with the ordered-start configuration
at $\beta=0.46$ and $m_q=0.006$, shown in
figure~\ref{fig:scalar_em_b046}.  We can see that although the
expectation value increases with increasing $\tau$, the error
increases at the same time.

\begin{figure}
  \vspace{-1.5ex}
  \centering
  \subfloat[Ordered-start, 440 configurations in total]
  {\includegraphics[width=0.49\textwidth]{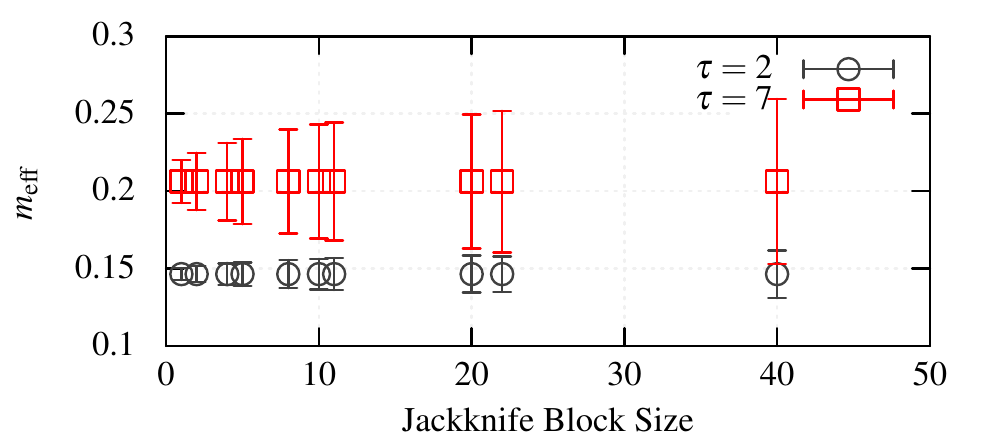}}
  \subfloat[Disordered-start, 320 configurations in total]
  {\includegraphics[width=0.49\textwidth]{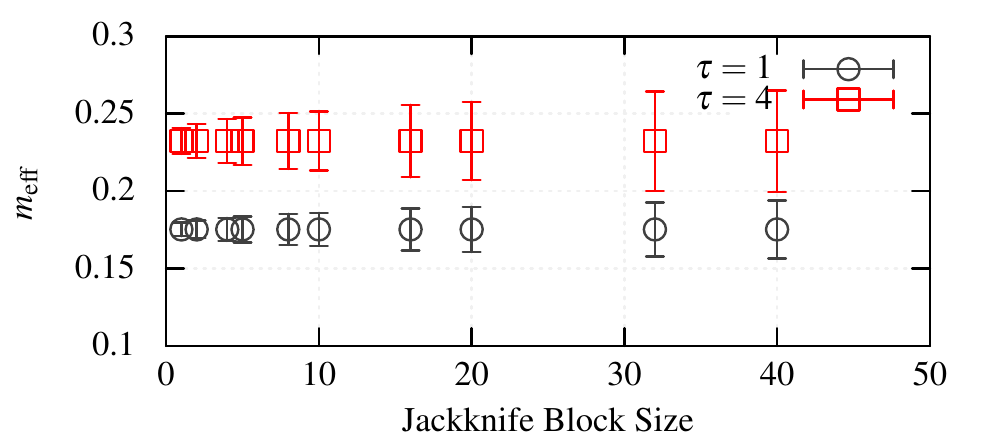}}
  \caption{\label{fig:scalar_em_error}Error of the effective masses of
    the scalar singlet meson versus the jackknife block size at
    $\beta=0.4626$ and $m_q=0.008$.}
\end{figure}

The errors of $m_\text{eff}$ are underestimated because of the long
autocorrelation time.  This is expected as the system approaches the
second order critical point.  As shown in
figure~\ref{fig:scalar_em_error}, the error bar grows as one blocks
the data in larger bins.  It is important to keep in mind that our
errors for the scalar single meson may be underestimated.

\begin{figure}
  \vspace{-3ex}
  \begin{minipage}{.45\textwidth}
    \centering
    \begin{tabular}[c]{llcl}
    $\beta$        & $m_q$     & $\tau$ fit range & $m_{\sigma^s}$ \\
    \hline\hline
    \mr{2}{0.46}   & 0.006$^O$ & $2\sim7$         & 0.295(30)      \\
                   & 0.006$^D$ & $1\sim4$         & 0.361(34)      \\
    \hline
    \mr{2}{0.4626} & 0.008$^O$ & $2\sim7$         & 0.160(14)      \\
                   & 0.008$^D$ & $1\sim4$         & 0.183(11)      \\
    \end{tabular}
  \end{minipage}
  \hfill
  \begin{minipage}{.5\textwidth}
    \centering
    \includegraphics[width=\textwidth]{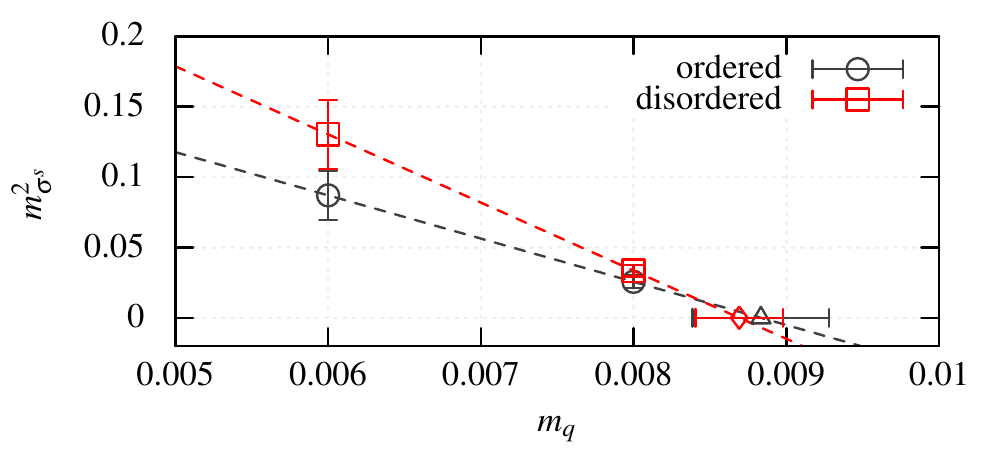}
  \end{minipage}
  \caption{\label{fig:scalar_singlet}Masses of the scalar singlet
    meson at the first order bulk transition, and the extrapolations
    to the critical point.}
  \vspace{-1ex}
\end{figure}

The mass of the scalar singlet meson was measured via fitting the
propagators.  The table at the left side of
figure~\ref{fig:scalar_singlet} shows the fitting range and the
results.  The statistical error quoted here were estimated using the
jackknife method with a block size of 8 configurations.  It is very
exciting to see that the scalar singlet meson becomes really light at
the point of the bulk transition at $\beta=0.4626$ and $m_q=0.008$,
since it is closer to the second order critical point.  If we apply
the scaling of universality class as $m_{\sigma^s} \sim C
(m_q-m_q^C)^{1/2}$, shown in figure~\ref{fig:scalar_singlet}, we can
estimate the critical quark mass at the end point to be at
$m_q^C=0.00883(45)$ from ordered-start configurations, and
$m_q^C=0.00869(29)$ from disordered-start configurations.

\vspace{-1ex}
\section{Conclusion}
\vspace{-2ex}

We have found a first order bulk transition with 12 light quark
flavors, using the DBW2 gauge and naive staggered action.  All the
hadronic observables are consistent with the system having
spontaneously broken chiral symmetry and asymptotic freedom on both
sides of the bulk transition.

The scalar singlet meson becomes lighter as the system approaches the
second order critical point (the end point of the first order bulk
transition), and is expected to have vanishing mass as other masses or
hadronic scales remain finite at the critical point.  This suggests
that the continuum limit of the system at this critical point is a
theory of scalars and hence one expects it to be a free theory, from
the general arguments of the triviality of scalar field theories.

The scaling behavior of the hadronic observables is expected to be
altered near this critical point, as opposed to the continuum limit
governed by the Gaussian fixed point at $\beta=\infty$.  Care must be
exercised to do any scaling test.  In particular, measuring the sigma
mass while approaching the conventional continuum limit should
indicate how much influence this critical endpoint has at a particular
point in parameter space.

The running of the gauge coupling directly measured on the lattice is
expected to be altered by this critical point, as it potentially
deviates from the behavior in the weak coupling limit.  More study is
required to quantify this effect.

Our calculations were done on the QCDOC and NY Blue at BNL.  X.-Y. Jin
wants to express the sincere gratitude especially toward all the
brilliant minds behind the QCDOC project.  This research utilized
resources at the New York Center for Computational Sciences at Stony
Brook University/Brookhaven National Laboratory which is supported by
the U.S. Department of Energy under Contract No. DE-FG02-92ER40699 and
by the State of New York.

\bibliographystyle{JHEP-2}
\bibliography{ref}

\end{document}